\documentclass[aps,prl,reprint,twocolumn,superscriptaddress,floatfix,nofootinbib,longbibliography]{revtex4-1}
\usepackage{graphicx,amsmath,amsfonts,amssymb,amsthm,xr}
\usepackage{epsfig,amsmath,amssymb,color,dsfont,upgreek,physics}
\usepackage{mathrsfs}
\usepackage{mathtools}
\usepackage{bbold}
\usepackage{comment}

\usepackage[bookmarks=true,colorlinks,linkcolor=OrangeRed,urlcolor=NavyBlue,citecolor=RoyalBlue]{hyperref}
\usepackage[dvipsnames]{xcolor}

\definecolor{mygold}{rgb}{0.93,0.69,0.13}

\definecolor{mypurple}{rgb}{0.49,0.18,0.56}

\definecolor{mygreen}{rgb}{0,0.5,0}

\definecolor{mygreen}{rgb}{0,0.5,0}
\definecolor{myred}{rgb}{0.7,0,0}
\usepackage{float}
\graphicspath{{./Figures/}}
\begin{document}

\title{Weak Ergodicity Breaking in the Schwinger Model}
\author{Jean-Yves Desaules}
\affiliation{School of Physics and Astronomy, University of Leeds, Leeds LS2 9JT, UK}
\author{Debasish Banerjee}
\affiliation{Theory Division, Saha Institute of Nuclear Physics, 1/AF Bidhan Nagar, Kolkata 700064, India}
\affiliation{Homi Bhabha National Institute, Training School Complex, Anushaktinagar, Mumbai 400094,India}
\author{Ana Hudomal}
\affiliation{School of Physics and Astronomy, University of Leeds, Leeds LS2 9JT, UK}
\affiliation{Institute of Physics Belgrade, University of Belgrade, 11080 Belgrade, Serbia}
\author{Zlatko Papi\'c}
\affiliation{School of Physics and Astronomy, University of Leeds, Leeds LS2 9JT, UK}
\author{Arnab Sen}
\affiliation{School of Physical Sciences, Indian Association for the Cultivation of Science, Kolkata 700032,
India}
\author{Jad C.~Halimeh}
\email{jad.halimeh@physik.lmu.de}
\affiliation{Department of Physics and Arnold Sommerfeld Center for Theoretical Physics (ASC), Ludwig-Maximilians-Universit\"at M\"unchen, Theresienstra\ss e 37, D-80333 M\"unchen, Germany}
\affiliation{Munich Center for Quantum Science and Technology (MCQST), Schellingstra\ss e 4, D-80799 M\"unchen, Germany}

\begin{abstract}
    As a paradigm of weak ergodicity breaking in disorder-free nonintegrable models, quantum many-body scars (QMBS) can offer deep insights into the thermalization dynamics of gauge theories. Having been first discovered in a spin-$1/2$ quantum link formulation of the Schwinger model, it is a fundamental question as to whether QMBS persist for $S>1/2$ since such theories converge to the lattice Schwinger model in the large-$S$ limit, which is the appropriate version of lattice QED in one spatial dimension. In this work, we address this question by exploring QMBS in spin-$S$ $\mathrm{U}(1)$ quantum link models (QLMs) with staggered fermions.
    We find that QMBS persist at $S>1/2$, with the resonant scarring regime, which occurs for a zero-mass quench, arising from simple high-energy gauge-invariant initial states. We furthermore find evidence of detuned scarring regimes, which occur for finite-mass quenches starting in the physical vacua and the charge-proliferated state. Our results conclusively show that QMBS exist in a wide class of lattice gauge theories in one spatial dimension represented by spin-$S$ QLMs coupled to dynamical fermions.
\end{abstract}
\date{\today}
\maketitle

\begin{figure}[t!]
\centering
\includegraphics[width=\linewidth]{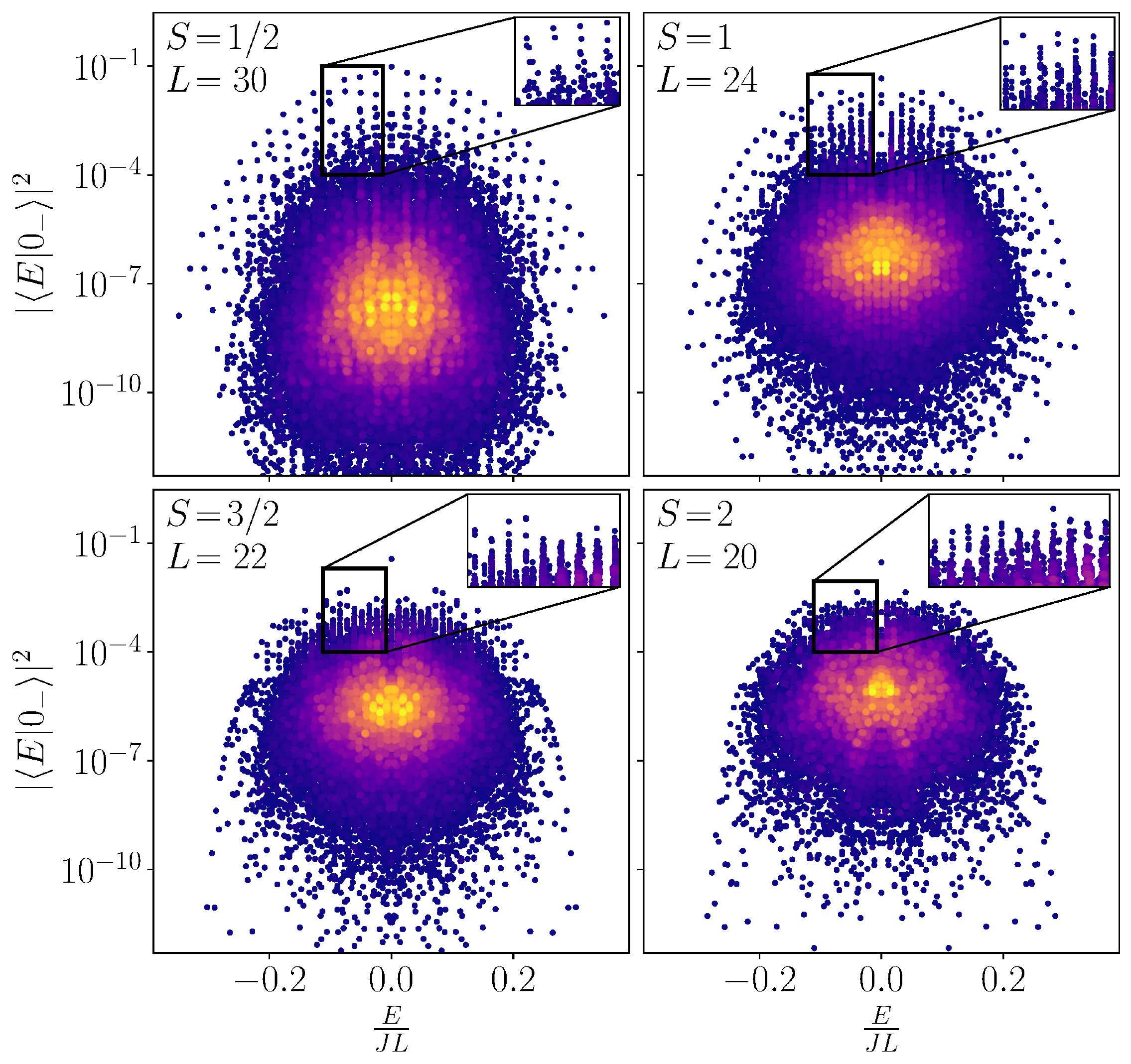}
\caption{(Color online). Quantum many-body scars in the spin-$S$ $\mathrm{U}(1)$ QLM: Overlap of the \textit{extreme vacuum} $\ket{0_-}$ with the eigenstates of the quench spin-$S$ $\mathrm{U}(1)$ QLM Hamiltonian~\eqref{eq:H} at $\mu=g=0$. At all considered values of $S$, distinctive towers of eigenstates arise that are equally spaced in energy (see insets). These are a hallmark of quantum many-body scars. These results are obtained from exact diagonalization calculations, where $L$ denotes the number of matter sites, and periodic boundary conditions are employed.}
\label{fig:LGT_olap}
\end{figure}

\textbf{\textit{Introduction.---}}Quantum many-body scars (QMBS) form an intriguing paradigm of ergodicity breaking in interacting systems that are typically expected to thermalize due to their nonintegrability and spatial homogeneity~\cite{Bernien2017,Moudgalya2018,Zhao2020,Zhao2021,Bluvstein2021,Jepsen2021,Su2022,zhang2022}. QMBS comprise eigenstates of low entanglement entropy~\cite{BernevigEnt,lin2018exact}, many of which reside in the middle of the spectrum, and are often separated roughly equally in energy~\cite{Turner2018,Iadecola2019_2}. 
These eigenstates are nonthermal, forming a ``cold'' subspace that is weakly connected to the rest of the Hilbert space. Consequently, quenches starting in initial states with high overlap with these nonthermal states (see, e.g., Fig.~\ref{fig:LGT_olap}) will not show typical thermalization, but instead exhibit long-lived coherent dynamics, persisting significantly beyond local relaxation times~\cite{wenwei18TDVPscar,Turner2018b}. 
This behavior is of particular interest to fundamental investigations of the eigenstate thermalization hypothesis (ETH) \cite{Srednicki1994,Deutsch1991,Rigol_2008,Eisert2015,Rigol_review,Deutsch_review}, as it has been linked to novel mechanisms for avoiding thermalization in closed quantum systems based on spectrum generating algebras~\cite{MotrunichTowers, MoudgalyaHubbard,Dea2020,Pakrouski2020} and embedding of nonthermal eigenstates~\cite{ShiraishiMori} (see recent reviews~\cite{Serbyn2020,MoudgalyaReview}). Moreover, given that many of these QMBS constitute high or even infinite-temperature states, when the system is initially prepared in them it will not dephase its information, which is pertinent to applications in quantum memory and information processing~\cite{Omran2019,Dooley2021,Desaules2021}.

QMBS are also relevant to gauge theories~\cite{Brenes2018,Surace2020,Banerjee2021,aramthottil2022scar,biswas2022scars}, which describe the interactions of elementary particles  mediated by gauge bosons through an extensive set of local constraints~\cite{Weinberg_book,Gattringer_book,Zee_book}. A paradigmatic example of the latter is Gauss's law in quantum electrodynamics (QED), where the distribution of charged matter strictly specifies the allowed configurations of the surrounding electromagnetic field~\cite{feynman1962quantum}. Recently, a concerted experimental effort has emerged for the implementation of gauge theories in synthetic quantum matter (SQM) devices \cite{Martinez2016,Muschik2017,Klco2018,Keesling2019, Kokail2019,Goerg2019,Schweizer2019,Mil2020,Klco2020,Yang2020,Zhou2021}. This has been facilitated in large part due to the great progress achieved in the precision and control of SQM setups \cite{Bloch2008,Bakr2009}, making the quantum simulation of gauge theories a realistic endeavor \cite{Wiese_review,Pasquans_review,Alexeev_review,aidelsburger2021cold,zohar2021quantum,klco2021standard,homeier2020mathbbz2}. Due to the complexity involved in these experiments, the implementations often focus on quantum link formulations of gauge theories, where spin-$S$ operators model the gauge fields, which in QED span an infinite-dimensional Hilbert space \cite{Chandrasekharan1997}. This has allowed the first large-scale realization of the spin-$1/2$ $\mathrm{U}(1)$ quantum link model (QLM) in $(1+1)-$D using ultracold atoms \cite{Yang2020,Zhou2021}.

The first experimental observation of QMBS was achieved  in a Rydberg-atom setup~\cite{Bernien2017} that maps to the spin-$1/2$ $\mathrm{U}(1)$ QLM~\cite{Surace2020}. It remains an open question whether QMBS persist at larger link spin lengths ($S>1/2$) in the QLM formulation of QED, and, if they do, what their form will be. In this Letter, we show that QMBS survive at any value of $S$. For a zero-mass quench, QMBS arise when the system is prepared in the \textit{extreme vacua} of the spin-$S$ $\mathrm{U}(1)$ QLM, which are the most highly excited vacuum states of lattice QED. Furthermore, we find that preparing the system in the physical (least excited) vacua or in the charge-proliferated state can still lead to \textit{detuned} scarring behavior for certain massive quenches, similar to the case of $S=1/2$ that has recently been demonstrated experimentally in a tilted Bose--Hubbard optical lattice \cite{Su2022}. Our results indicate that QMBS may persist toward the limit of lattice QED, $S\to\infty$.

\textbf{\textit{$\mathrm{U}(1)$ quantum link model.---}}The Schwinger model, or QED in $(1 + 1)$ dimensions, is possibly the simplest gauge theory with dynamical matter and that shows 
nontrivial phenomena like confinement \cite{Schwinger1962,Coleman1975}. A possible discrete version of the Schwinger Hamiltonian on a lattice is provided by 
the Kogut--Susskind formulation, which is also reached in the large-$S$ limit of the QLMs being studied here. The $(1+1)-$D spin-$S$ $\mathrm{U}(1)$ QLM is given by the Hamiltonian \cite{Banerjee2012,kasper2017}
\begin{align}\nonumber
    \hat{H}=&\sum_{j=1}^L\bigg[\frac{J}{2a\sqrt{S(S+1)}}\big(\hat{\sigma}_j^+\hat{s}^+_{j,j+1}\hat{\sigma}^-_{j+1}+\text{H.c.}\big)\\\label{eq:H}
    &+\frac{\mu}{2}(-1)^j\hat{\sigma}^z_j+\frac{g^2a}{2}\big(\hat{s}^z_{j,j+1}\big)^2\bigg].
\end{align}
Here, $J=1$ sets the energy scale, $\mu$ is the fermionic mass, and $g^2$ is the electric-field coupling strength. Throughout this work, we will set the lattice spacing to $a=1$ and employ periodic boundary conditions, with $L$ denoting the number of lattice sites. The matter field on site $j$ is represented by the Pauli operator $\hat{\sigma}^z_j$, and the electric (gauge) field at the link between sites $j$ and $j+1$ is represented by the spin-$S$ operator $\hat{s}^{z(+)}_{j,j+1}$. The generator of the $\mathrm{U}(1)$ gauge symmetry of Hamiltonian~\eqref{eq:H} is 
\begin{align}\label{eq:Gj}
    \hat{G}_j=\frac{\hat{\sigma}^z_j+(-1)^j}{2}+\hat{s}^z_{j-1,j}-\hat{s}^z_{j,j+1},
\end{align}
which can be interpreted as a discretized version of Gauss's law relating the matter occupation on site $j$ to the electric-field configuration on its neighboring links. We will work in the \textit{physical} sector of Gauss's law: $\hat{G}_j\ket{\phi}=0,\,\forall j$.

Due to the gauge symmetry imposed by the generator~\eqref{eq:Gj}, one can integrate out the matter fields in the Hamiltonian~\eqref{eq:H}, resulting in a constrained spin system. For $S=1/2$, this corresponds to the PXP model \cite{Surace2020}. For larger $S$, the resulting model differs from generalizations of the PXP model already explored in the literature \cite{wenwei18TDVPscar,Mukherjee2021}. In the joint submission~\cite{Desaules2022prominent}, we derive the relevant constrained spin-$S$ model corresponding to the spin-$S$ $\mathrm{U}(1)$ QLM for any value of $S$ \cite{Zache2021achieving}. Exact diagonalization (ED) techniques resolving the translation and spatial-inversion symmetries have been employed to study the eigenstates of these models. Time-evolution results are obtained either directly from the ED results or by time-evolving the initial state using sparse matrix exponential techniques.

\textbf{\textit{Resonant scarring.---}}
The \textit{physical vacuum} of $\hat{H}$ is its ground state at $\mu\to\infty$ and $g^2>0$. In the case of half-integer $S$, there are two doubly degenerate physical vacua. These can be defined on a two-site two-link unit cell using as quantum numbers the eigenvalues $\sigma^z_j$ and $s^z_{j,j+1}$ of the matter and electric field operators $\hat{\sigma}^z_j$ and $\hat{s}^z_{j,j+1}$, respectively, explicitly reading $\ket{\sigma^z_1,s^z_{1,2},\sigma^z_2,s^z_{2,3}}=\ket{+1,\pm1/2,-1,\pm1/2}$. For integer $S$, the physical vacuum is nondegenerate, and reads $\ket{\sigma^z_1,s^z_{1,2},\sigma^z_2,s^z_{2,3}}=\ket{+1,0,-1,0}$. Henceforth, we will denote $\ket{0_+}=\ket{+1,+1/2,-1,+1/2}$ for half-integer $S$ and $\ket{0_+}=\ket{+1,0,-1,0}$ for integer $S$, with the subscript denoting the sign of $g^2$.

On the other hand, the \textit{extreme vacua} of $\hat{H}$ are high-energy states that can be realized as doubly degenerate ground states of Eq.~\eqref{eq:H} at $\mu\to\infty$ and $g^2<0$: $\ket{\sigma^z_1,s^z_{1,2},\sigma^z_2,s^z_{2,3}}=\ket{+1,\pm S,-1,\pm S}$. Henceforth, we will denote $\ket{0_-}=\ket{+1,+S,-1,+S}$, with the subscript again indicating the sign of $g^2$.

We will further consider the charge-proliferated state, which is the ground state of Eq.~\eqref{eq:H} at $\mu\to-\infty$ and $g^2>0$. For half-integer $S$, it is nondegenerate and reads $\ket{\mathrm{CP}}=\ket{-1,-1/2,+1,+1/2}$. For integer $S$, we obtain two doubly degenerate ground states: $\ket{\mathrm{CP}}=\ket{-1,-1,+1,0}$ and $\ket{-1,0,+1,+1}$.

\begin{figure}[t!]
\centering
\includegraphics[width=\linewidth]{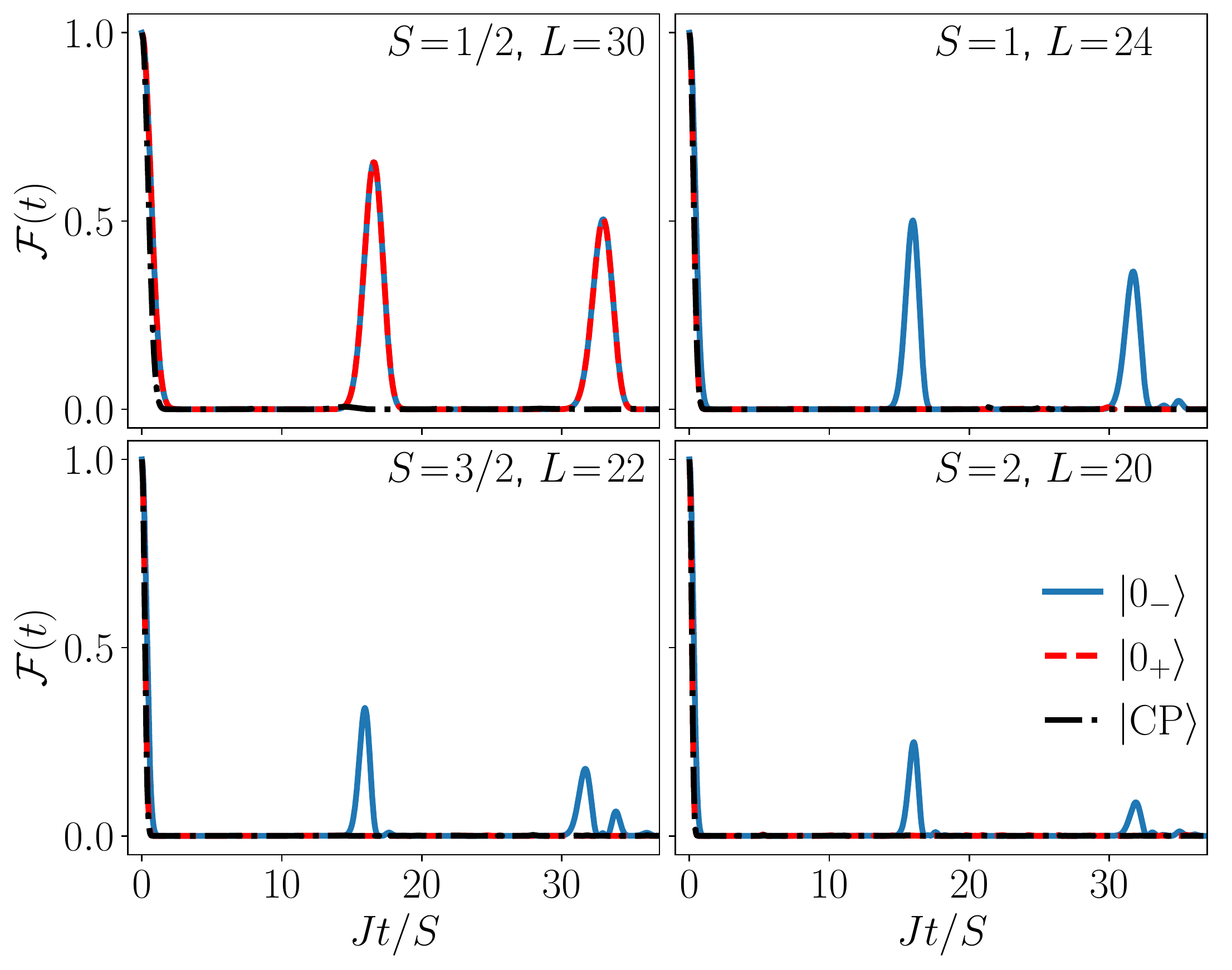}
\caption{(Color online). Dynamics of the fidelity in the wake of a quench by the spin-$S$ $\mathrm{U}(1)$ QLM~\eqref{eq:H} at $\mu=g=0$ starting in either the extreme vacuum $\ket{0_-}$ (solid blue curve), the physical vacuum $\ket{0_+}$ (dashed red curve), or the charge-proliferated state $\ket{\text{CP}}$ (dash-dotted black curve). The qualitative conclusion is that regardless of the value of $S$, prominent revivals appear in the fidelity dynamics only when the initial state is the extreme vacuum, whereas for other states the dynamics is thermal. Note that for $S=1/2$, the physical and extreme vacua are identical.}
\label{fig:fidelity}
\end{figure}

With respect to the eigenstates of Hamiltonian~\eqref{eq:H} at $\mu=g=0$, we find through ED that only $\ket{0_-}$ exhibits the overlap behavior indicative of scarring for general $S$; see Fig.~\ref{fig:LGT_olap}. Just as in the known case of $S=1/2$, we also see at other values of $S$ signatures of $2SL+1$ towers equally spaced in energy (see insets), exhibiting large overlap with $\ket{0_-}$, particularly in the middle of the spectrum. The overlap of the top band of states can be further enhanced by considering a truncated version of the QED gauge field~\cite{Desaules2022prominent}. Note how for all values of $S$ that we consider, there is a prominent zero-energy mode with the largest overlap. The presence of these eigenstates is evidence of weak ergodicity breaking in the model. Due to the scaling term $1/\sqrt{S(S+1)}$, the ground-state energy $E_0$ is approximately independent of $S$ at $\mu=g^2=0$, and we find numerically that $E_0\approx -0.32 L$. As the spectrum is symmetric around zero, we can use this along with the number of towers to get the approximate energy spacing between towers as $\Delta E\approx -2E_0/(2SL)\approx0.32/S$.
We note that the various approximation schemes for scarred eigenstates in the PXP model also show good results for larger $S$ \cite{Desaules2022prominent}.

\begin{figure}[t!]
\centering
\includegraphics[width=0.93\linewidth]{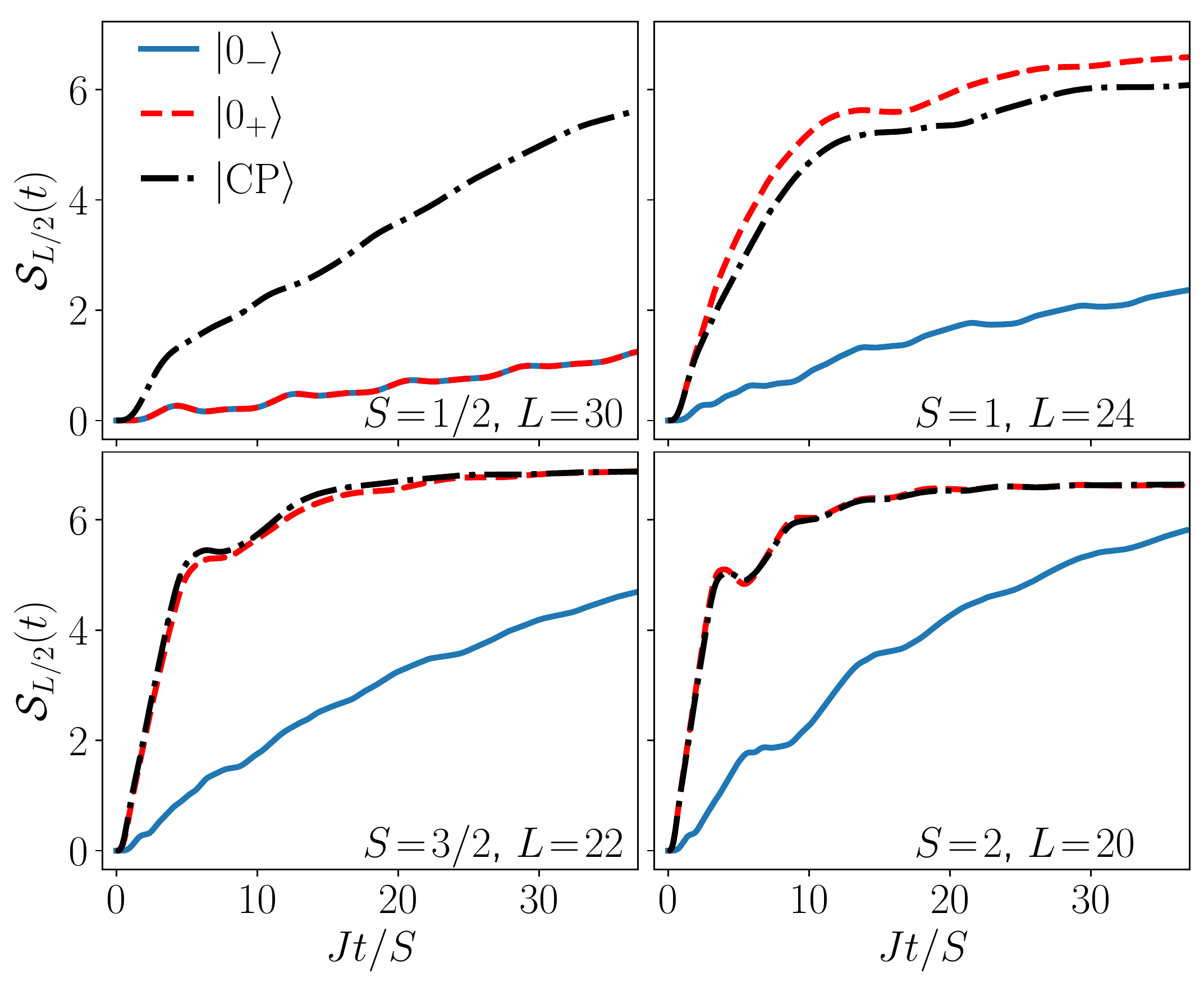}
\caption{(Color online). Dynamics of the mid-chain entanglement entropy for the same quench and initial states considered in Fig.~\ref{fig:fidelity}. For all considered values of the link spin length $S$, an anomalously low and slowly growing entanglement entropy arises when the initial state is the extreme vacuum. Preparing the system in the charge-proliferated state or the physical vacuum leads to a fast growth in the entanglement entropy except for the case of $S=1/2$, where the physical and extreme vacua are the same.}
\label{fig:LGT_S_evol}
\end{figure}

The presence of scarred eigenstates can be detected via the dynamics of the fidelity, $\mathcal{F}(t)=\lvert\braket{\psi(0)}{\psi(t)}\rvert^2$ with $\ket{\psi(t)}=e^{-i\hat{H}t}\ket{\psi(0)}$ and $\ket{\psi(0)}$ the initial state. In the case of $S=1/2$, $\ket{0_+}=\ket{0_-}$, as the last term of Eq.~\eqref{eq:H} is an inconsequential energy constant since $(\hat{s}^z_{j,j+1})^2=\mathds{1}$. Quenching this vacuum state with $\hat{H}$ at $\mu=0$ is known to lead to scarring behavior for $S=1/2$~\cite{Bernien2017, Surace2020}, and this is exhibited in the revivals of the fidelity, shown in the top left panel of Fig.~\ref{fig:fidelity}. For comparison, we have included the fidelity dynamics for $\ket{\psi(0)}=\ket{\mathrm{CP}}$, which shows no revivals, in agreement with what is established in the literature for this quench when $S=1/2$ \cite{Serbyn2020}.

However, once the link spin length is $S>1/2$, we find that the fidelity dynamics exhibits revivals only when the system is initialized in the extreme vacuum, $\ket{\psi(0)}=\ket{0_-}$, whereas neither the physical vacuum $\ket{0_+}$ nor the charge-proliferated state $\ket{\text{CP}}$ give rise to scarring behavior; see Fig.~\ref{fig:fidelity} for $S=1,\,3/2,\,2$. We have checked that the other vacua $\ket{\sigma^z_1,s^z_{1,2},\sigma^z_2,s^z_{2,3}}=\ket{+1,\pm M,-1,\pm M}$ with $1/2<M<S$ and higher-energy charge-proliferated states are also not scarred states~\cite{Desaules2022prominent}. From the previous estimate of the energies of scarred towers, we expect the revival period to be $T\approx 6.25\pi S$. However, in practice the energy spacing between towers varies along the spectrum. So the relevant energy spacing to consider is the one near $E=0$, where the scarred states have the higher overlap with $\ket{0_-}$. This provides an estimate of $T\approx 5.13\pi S$, which much more accurately agrees with the numerical data.

\begin{figure}[t!]
\centering
\includegraphics[width=0.93\linewidth]{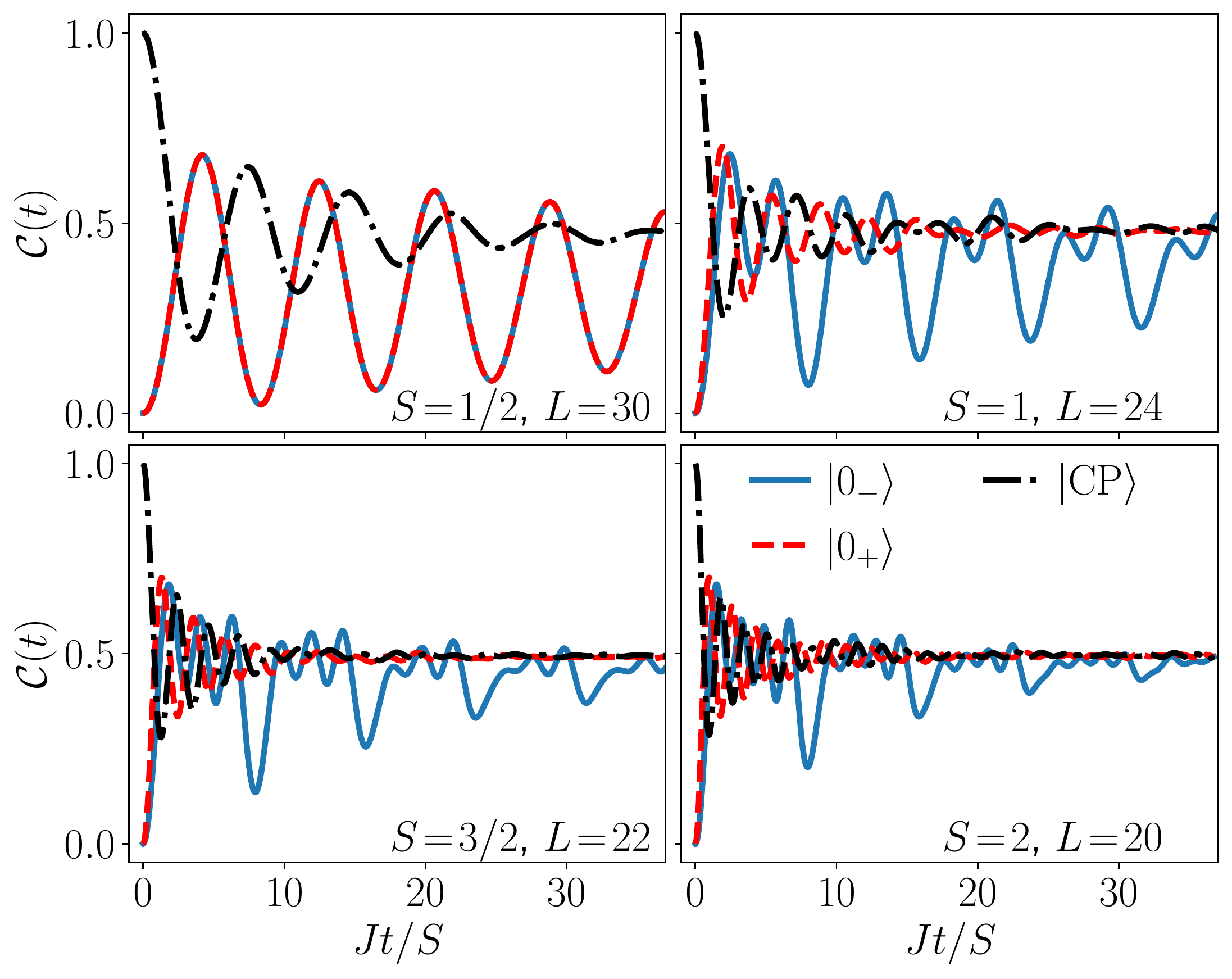}
\caption{(Color online). Dynamics of the chiral condensate~\eqref{eq:CC} in the wake of the same quench and initial states considered in Fig.~\ref{fig:fidelity}. As is characteristic of scarred dynamics, local observables show persistent oscillations up to all accessible times when the initial state is prepared in the extreme vacuum. In all other cases, the chiral condensate quickly thermalizes. We note again the exception for the case of $S=1/2$, where the physical vacuum is itself the extreme vacuum.}
\label{fig:LGT_M_evol}
\end{figure}

We explore the effect of scarring on the dynamics of the mid-chain entanglement entropy $\mathcal{S}_{L/2}(t)$, shown in Fig.~\ref{fig:LGT_S_evol} for $S=1/2$ to $2$. In all cases, starting in the extreme vacuum leads to an anomalously low $\mathcal{S}_{L/2}(t)$ exhibiting significantly slower growth, whereas preparing the system in the charge-proliferated state leads to a rapid increase in the entanglement entropy. Except for the case of $S=1/2$ where the extreme and physical vacua are the same, starting in the physical vacuum leads to behavior qualitatively similar to that of the charge-proliferated state, with a rapid growth in $\mathcal{S}_{L/2}(t)$. These findings are consistent with nonthermal scarred dynamics only when the initial state is prepared in the extreme vacuum. 

Of great experimental interest in the observation of scarring behavior are local observables, which are expected to show persistent oscillations up to times sufficiently longer than the fastest timescales of the system \cite{Bernien2017}. In the following, we explore this by computing the dynamics of the chiral condensate
\begin{align}\label{eq:CC}
\mathcal{C}(t)=\frac{1}{2}+\frac{1}{2L}\sum_{j=1}^L(-1)^j\bra{\psi(t)}\hat{\sigma}^z_j\ket{\psi(t)},   
\end{align}
which is a measure of how strongly the chiral symmetry corresponding to fermions in the
theory is (spontaneously) broken by the dynamics of the theory. The results shown in Fig.~\ref{fig:LGT_M_evol} for $S=1/2$ to $2$ show strikingly nonthermal behavior for quenches starting in $\ket{0_-}$. Indeed, over the timescales we simulate there are persistent oscillations in the signal and no equilibration, in agreement with experimental results for the spin-$1/2$ $\mathrm{U}(1)$ QLM \cite{Bernien2017,Su2022}. The oscillations in $\mathcal{C}(t)$ have twice the frequency of the fidelity revivals. This happens when the state comes back to itself, but also when there is state transfer to the other extreme vacuum. As $S$ is increased, we also see more oscillations in-between the large ones. These consist of an alternation of vacua and charge-proliferated states with different value of field \cite{Desaules2022prominent}. On the other hand, systems prepared in $\ket{0_+}$ (for $S>1/2$) or $\ket{\text{CP}}$ show relatively fast thermalization, where oscillations are quickly suppressed in the dynamics, consistent with the ETH.

As such, we have demonstrated that for a quench at $\mu=g=0$, the spin-$S$ $\mathrm{U}(1)$ QLM~\eqref{eq:H} exhibits scarring behavior when the system is initially prepared in an \textit{extreme vacuum}. 
The underlying scarring mechanism is precession of a ``large" spin of magnitude $SL$~\cite{Choi2018}.   
The relation between scarring and the Hilbert space constraint can be firmed up by studying the structure of the adjacency graph of the Hamiltonian \cite{Desaules2022prominent}. These extreme vacua are product states that can be naturally explored in SQM experiments \cite{Yang2020,Zhou2021} even though they are otherwise inaccessible in lattice QED.

\textbf{\textit{Detuned scarring.---}}In a recent study \cite{Su2022}, it has been shown theoretically and demonstrated experimentally that there are scarring regimes beyond the resonant one discussed above. This was demonstrated in the spin-$1/2$ $\mathrm{U}(1)$ QLM by starting in the charge-proliferated state and performing a quench at finite mass  (detuning).

\begin{figure}[t!]
\centering
\includegraphics[width=\linewidth]{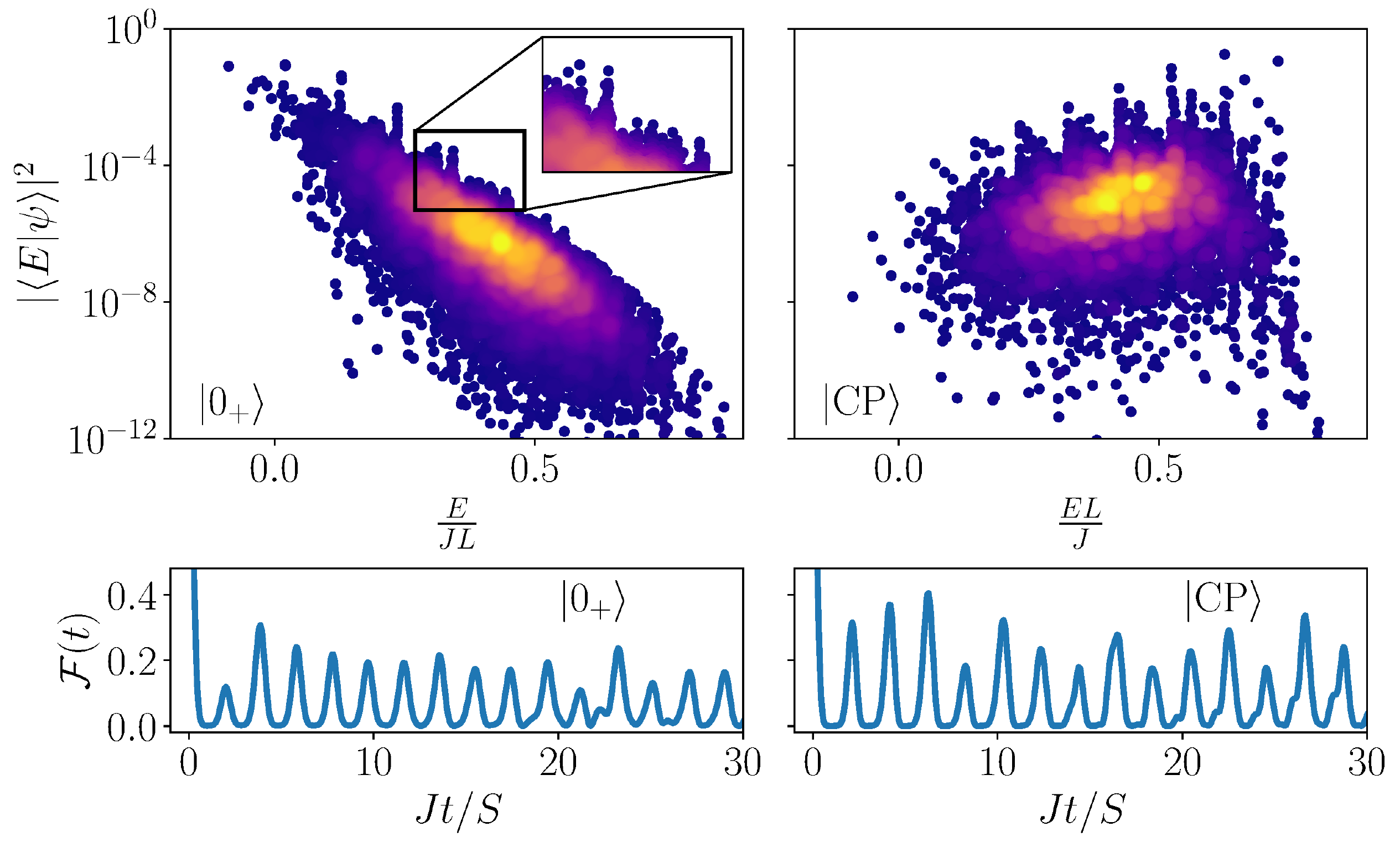}
\caption{(Color online). The spin-$S$ $\mathrm{U}(1)$ QLM also exhibits detuned scarring \cite{Su2022}, where the physical vacuum $\ket{0_+}$ and the charge-proliferated state $\ket{\text{CP}}$ can lead to scarred dynamics when quenched by Hamiltonian~\eqref{eq:H} at small nonzero values of $\mu$ and $g^2$. Here, we set $\mu=0.486J$ and $g^2=0.6J$, and the overlap of each of $\ket{0_+}$ and $\ket{\text{CP}}$ with the corresponding eigenstates of Hamiltonian~\eqref{eq:H} are shown for $L=20$ and $S=3/2$. Towers of eigenstates equally spaced in energy emerge, indicating the presence of QMBS. We confirm this picture by calculating the dynamics of the fidelity in the wake of this quench. For all accessible evolution times, we find consistent revivals in the fidelity, strongly indicative of scarred dynamics.}
\label{fig:QLM_unconv}
\end{figure}

Motivated by the question as to whether QMBS persist in lattice QED for physically relevant initial states, we explore these \textit{detuned} scarring regimes in the spin-$3/2$ $\mathrm{U}(1)$ QLM starting in either the physical vacuum $\ket{0_+}$ or the charge-proliferated state $\ket{\text{CP}}$. As shown in Fig.~\ref{fig:QLM_unconv}, the overlap of these initial states with the eigenstates of the quench Hamiltonian~\eqref{eq:H} at $\mu=0.486J$ and $g^2=0.6J$ shows distinctive towers equally spaced in energy, similar to the known case of $S=1/2$ \cite{Su2022}. Also displayed in Fig.~\ref{fig:QLM_unconv} is the fidelity dynamics for each of $\ket{0_+}$ and $\ket{\text{CP}}$ upon quenching them with this Hamiltonian, where we see persistent revivals up to all considered evolution times. We also arrive at a similar picture for other values of $S$, and in fact we find a wide range of values of $(\mu,g^2)$ over which scarring behavior emerges \cite{Desaules2022prominent}.

Given that $\ket{0_+}$ and $\ket{\text{CP}}$ are both physically relevant in lattice QED in one spatial dimension, and since the latter has been shown to be achieved at relatively small values of $S$ both in \cite{Buyens2017,Banuls2020,Zache2021achieving} and out of equilibrium \cite{Halimeh2021achieving}, our results suggest that QMBS may play a role for understanding dynamics in physically interesting regimes as well.

\textbf{\textit{Summary.---}}In conclusion, we have investigated QMBS in the paradigmatic spin-$S$ $\mathrm{U}(1)$ QLM, a staple of modern SQM experiments on lattice gauge theories. We have shown that the regime of resonant scarring for quenches at zero mass, prevalent in the literature in the case of $S=1/2$, is also present in the case of $S>1/2$ when the system is initially prepared in an \textit{extreme vacuum}, where the local electric field takes on its largest possible eigenvalue. This has been done by calculating in ED the overlap of the extreme vacuum with the eigenstates of the quench Hamiltonian, showing clear towers equally spaced in energy, which are a hallmark of QMBS. Furthermore, we have shown consistent revivals in the dynamics of the fidelity, an anomalously low and slowly growing mid-chain entanglement entropy, and persistent oscillations in the dynamics of the chiral condensate over all times. These extreme vacua are not physical as ground states in lattice QED yet are easily implementable product states exhibiting clear scarring behavior that can be explored in SQM experiments.

We have also uncovered detuned scarring in the spin-$S$ $\mathrm{U}(1)$ QLM arising from quenches at small nonzero mass and electric-field coupling, when the system is initially prepared in either the physical vacuum, where the local electric field takes on its lowest possible eigenvalue, or the charge-proliferated state. The overlap of these states with the eigenstates of the quench Hamiltonian exhibits distinctive towers at equal spacing in energy, and the corresponding fidelity dynamics shows consistent revivals over all accessible times. These initial states are physically relevant as low-energy states in lattice QED. Given that recent works have shown convergence to the latter limit in and out of equilibrium already at $S\gtrsim3/2$, our results suggest that this detuned scarring regime may already exist in lattice QED.

\begin{acknowledgments}
J.C.H.~and J.-Y.D.~are very grateful to Giuliano Giudici for insightful discussions and valuable comments. J.C.H.~acknowledges funding from the European Research Council (ERC) under the European Union’s Horizon 2020 research and innovation programm (Grant Agreement no 948141) — ERC Starting Grant SimUcQuam, and by the Deutsche Forschungsgemeinschaft (DFG, German Research Foundation) under Germany's Excellence Strategy -- EXC-2111 -- 390814868. We acknowledge support by EPSRC grants EP/R020612/1 (Z.P.) and EP/R513258/1 (J.-Y.D.). A.H.~and Z.P. acknowledge support by the Leverhulme Trust Research Leadership Award RL-2019-015. A.H.~acknowledges funding provided by the Institute of Physics Belgrade, through the grant by the Ministry of Education, Science, and Technological Development of the Republic of Serbia. Statement of compliance with EPSRC policy framework on research data: This publication is theoretical
work that does not require supporting research data.
\end{acknowledgments}

\bibliography{Schwinger_biblio.bib}

\end{document}